# Total reflection and transmission by epsilon-near-zero metamaterials with defects

*Yadong Xu and Huanyang Chen\**

*School of Physical Science and Technology, Soochow University, Suzhou 215006, China*

**Abstract:** In this work, we investigate wave transmission through an epsilon-near-zero metamaterial waveguide embedded with defects. We show that by adjusting the geometric sizes and material properties of the defects, total reflection and even transmission can be obtained, despite the impedance mismatch of epsilon-near-zero material with free space. Our work can greatly simplify the design of zero-index material waveguide applications by removing the dependence on permeability.

Recently, the manmade artificial materials denoted metamaterials [1, 2] have drawn extensive attentions due to their numerous novel applications [3, 4, 5]. Among them are double negative materials [1, 2, 6], single negative materials [7], epsilon-near-zero (ENZ) matematerials [8] and matched impedance zero-index materials (MIZIM) [9], *etc*. Zero-index materials (ZIM, including both ENZ and MIZIM), as a new type of metamaterials, have been explored both theoretically and experimentally to show a lot of intriguing properties [8, 9, 10, 11, 12, 13, 14, 15, 16, 17]. For example, Enoch *et al.* [8] showed that a ZIM can enhance the directive emission for an embedded source; Ziolkowski [9] studied the possibility of designing a MIZIM; Li *et al.* [10] proved that there is a zero-$\bar{n}$ gap inside the zero (volume) averaged refractive index material, which is distinct from the Bragg gaps; Silveirinha and Engheta [11-13] demonstrated an ENZ medium that can "squeeze" the electromagnetic (EM) waves in a narrow waveguide. Such a tunneling effect was later demonstrated in microwave experiments [15, 16]. Recently, Hao *et al.* [18] showed that a total reflection or transmission can be obtained by introducing perfect electric conductor (PEC) (or perfect magnetic conductor, PMC) defects inside the ZIM in a two dimensional (2D) waveguide structure. Nguyen *et al.* [19] found that similar effects can happen when dielectric defects are introduced into the MIZIM, which suggests an active control transmission and reflection by incorporating tunable refractive index materials. However, it is much more challenging to fabricate the MIZIM than the ENZ medium, because it is very difficult to engineer effective permittivity and permeability to be zero at the same time. In this letter, we will revisit the problem wave transmission in a similar waveguide structure with defects but replacing the MIZIM with the ENZ medium. Analytic expressions will be derived to see how total reflection and transmission can be achieved by adjusting the geometric sizes and material parameters of the embedded defects. Finite element numerical simulations will also be carried out to prove our theory.

---

\* kenyon@ust.hk

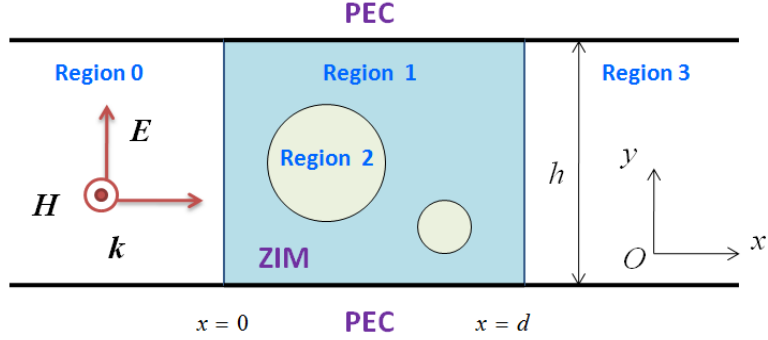

FIG. 1 (color online). The computation domain of the two dimensional waveguide structure with a ZIM (region 1). Region 0 and 3 are vacuum. Region 2 (the cylinders) are the embedded defects. The parallel black lines are PEC walls of waveguide. A TM mode is incident from left to right inside the waveguide.

In the beginning, we would like to consider a 2D waveguide structure, which consists of four regions as shown in FIG. 1. Region 0 and 3 are free spaces separated by a ZIM (region 1) with effective permittivity $\epsilon_1$ and permeability $\mu_1$. $N$ cylindrical defects (region 2) are embedded inside region 1, whose effective permittivities and permeabilities are $\epsilon_{2i}$ and $\mu_{2i}$, respectively (for the *ith* cylinder). For simplicity, we focus on the transverse magnetic (TM) polarization (the magnetic field **H** is polarized in z direction). The walls of the waveguide are set to be PECs (Similar results will be obtained for transverse electric (TE) mode where the walls of the waveguide are PMCs).

Suppose that a TM mode with $\mathbf{H}_{\text{in}} = \hat{z}\, H_{0i} e^{i(k_0 x - \omega t)}$ is incident from left to right inside the above waveguide, where $k_0 (= \omega/c)$ is the wave verctor and $\omega$ is the angular frequency. We will omit the time variation $e^{-i\omega t}$ throughout the following for convenience. The EM wave in each region follows the Ampére − Maxwell equation,

$$E_n = -\frac{1}{i\omega\epsilon_0\epsilon_n}\nabla \times H_n, \qquad (1)$$

where the integer $n$ signifies each region, $\epsilon_n$ is the relative permittivity of each region. The electromagnetic field in region 0 can be written as,

$$H_0 = \hat{z}\, H_{0i}(e^{ik_0 x} + \mathcal{R}e^{-ik_0 x}), \qquad (2)$$

and

$$E_0 = \frac{k_0}{\omega\epsilon_0}\hat{y}H_{0i}(e^{ik_0 x} - \mathcal{R}e^{-ik_0 x}), \qquad (3)$$

where $\mathcal{R}$ is the reflection coefficient. Likewise in region 3, we have,

$$H_3 = \hat{z}\, TH_{0i}e^{ik_0(x-d)}, \qquad (4)$$

$$E_3 = -\frac{1}{i\omega\epsilon_0}\nabla \times H_0 = \frac{k_0}{\omega\epsilon_0}\hat{y}\, TH_{0i}e^{ik_0(x-d)}, \qquad (5)$$

where $T$ is the transmission coefficient. For region 1, as its $\epsilon_1 \cong 0$, in order to have a finite value of the electric field [19], $\nabla \times H_1$ must vanish, *i.e.*, $H_1 = \hat{z}H_1$ with a constant value $H_1$. The boundary condition at the interface of region 0 and 1 gives that $H_{0i} + \mathcal{R}H_{0i} = H_1$, while another one at interface of region 1 and 3 is $TH_{0i} = H_1$. Therefore we obtain a simple relationship between the two transmission and reflection coefficients, $T = 1 + \mathcal{R}$. For region 2, the magnetic field inside each cylindrical defect obeys the Helmholtz equation,

$$\nabla^2 H_2 + k_0^2 \epsilon_{2i}\mu_{2i}H_2 = 0. \qquad (6)$$

We will ignore the magnetic coupling between the defects as in Ref. [19]. As $H_1$ is a constant, Dirichlet boundary conditions should be applied at the surface of each defect and lead to the magnetic field distribution as [19],

$$H_2 = \hat{z}\, H_1 \sum_{i=1}^{N} \frac{J_0(k_{2i}r_i)}{J_0(k_{2i}R_i)}, \qquad (7)$$

where $J_0$ is the zero-order Bessel function of the first kind; $k_{2i} = k_0\sqrt{\epsilon_{2i}\mu_{2i}}$ is the wave vector in each cylindrical defect; $R_i$ is the radius of each cylinder; $r_i$ is relative coordinate as that in Ref. [19]. The electric field inside the defects can also be obtained as [19],

$$E_2 = iH_1 \sum_{i=1}^{N} \frac{J_1(k_{2i}r_i)}{J_0(k_{2i}R_i)} \sqrt{\frac{\mu_{2i}}{\epsilon_{2i}}}\hat{\theta}_i, \qquad (8)$$

where $J_1$ is the *1st*-order Bessel function of the first kind, $\hat{\theta}_i$ is the azimuthal unit vector for the *ith* cylindrical defect. Using Maxwell-Faraday equation,

$$\oint E\, dl = -\int \frac{\partial B}{\partial t}ds, \qquad (9)$$

we can find out the transmission coefficient as,

$$T = \frac{1}{1 - \frac{ik_0\mu_1(S-S_d)}{2h} - \frac{i\pi}{h}\sum_{i=1}^{N}\left[\frac{R_i J_1(k_{2i}R_i)}{J_0(k_{2i}R_i)}\right]\sqrt{\frac{\mu_{2i}}{\epsilon_{2i}}}}, \qquad (10)$$

where $S = d \times h$ is total area of the region 1 and 2, $S_d = \sum_{i=1}^{N} \pi R_i^2$ is total sum of the areas of N cylindrical defects.

Let us now discuss the above transmission coefficient expression. (a), if $\mu_1$ vanishes, the ZIM becomes a matched impedance zero-index metamaterial (MIZIM), Eq. (10) will go back to the transmission coefficient formula in Ref. [19] (its equation (8)). (b), if any of the defects is PMC ($\mu_{2i} = -\infty$), $T = 0$, we can obtain similar results in Ref. [18]. (c), if $\mu_1$ is finite value and no defect exists in the region 1 (*i.e.*, $S_d = 0$), the third term of the denominator in Eq. (10) will disappear, thereby the transmission coefficient $T$ will reduce to,

$$T = \frac{1}{1 - \frac{ik_0 \mu_1 d}{2}}, \quad (11)$$

which is in accordance with the results in Ref. [14].

According to Equation (10), total reflection or transmission can also be achieved by embedding proper defects inside the ZIM. For example, if $J_0(k_{2i}R_i)$ is equal to zero, no matter what value of the permeability of ZIM is, there will be a total reflection. Here we choose $\mu_1 = 1$, such that ZIM is of a nonmagnetic response, *i.e.*, the ENZ medium which could be easily designed and fabricated. In order to verify our analysis, numerical simulations are performed by using finite element solver COMSOL MULTIPHYSICS. We set $d$=32mm, $h$=30mm. The frequency of the incoming TM wave is 10GHz. For simplicity, here we only consider one cylindrical defect with radius R=8mm. The permittivity and permeability of the defect are chosen to be $\epsilon_2 = 2.06$ and $\mu_2 = 1$ to satisfy $J_0(k_{2i}R_i) = 0$, which denotes a dielectric material.

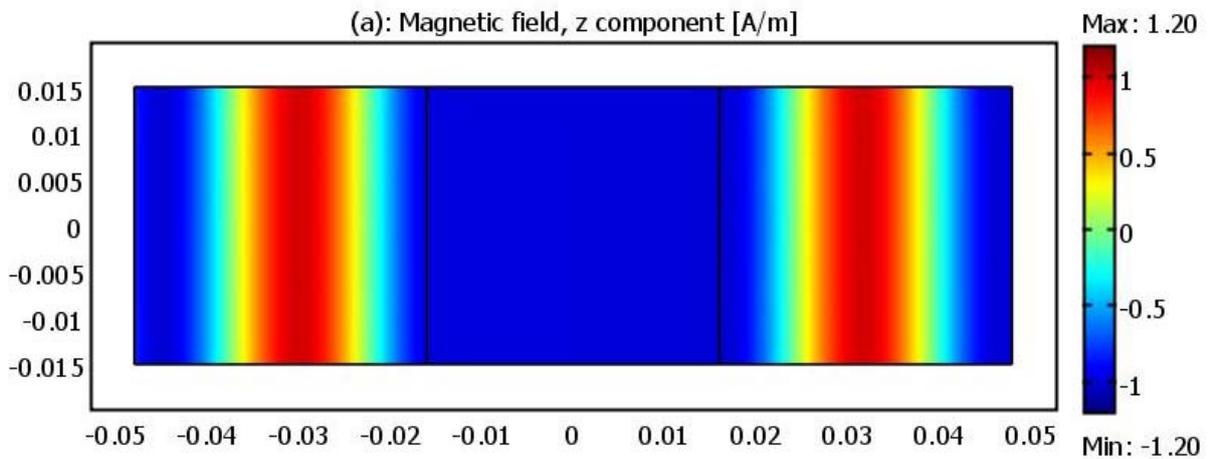

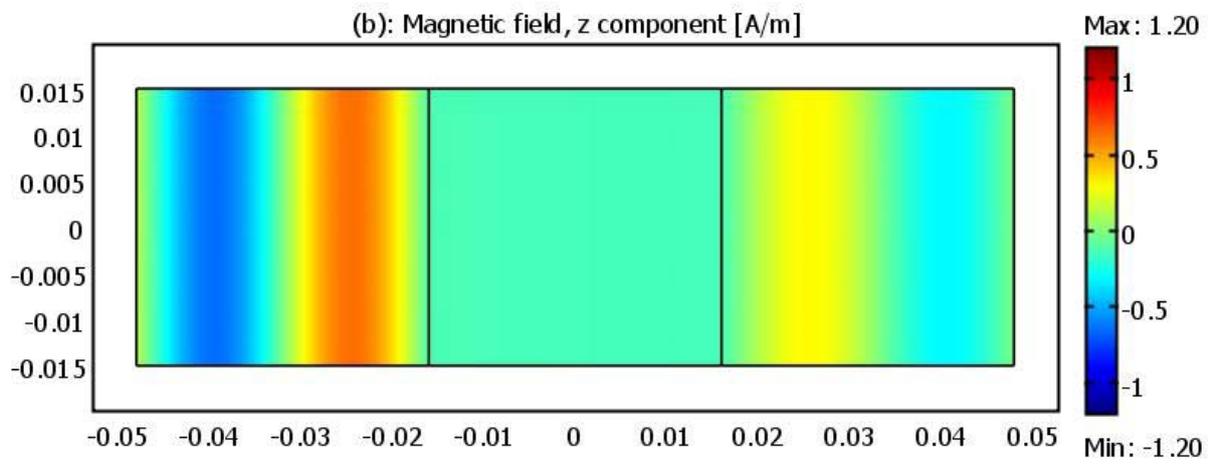

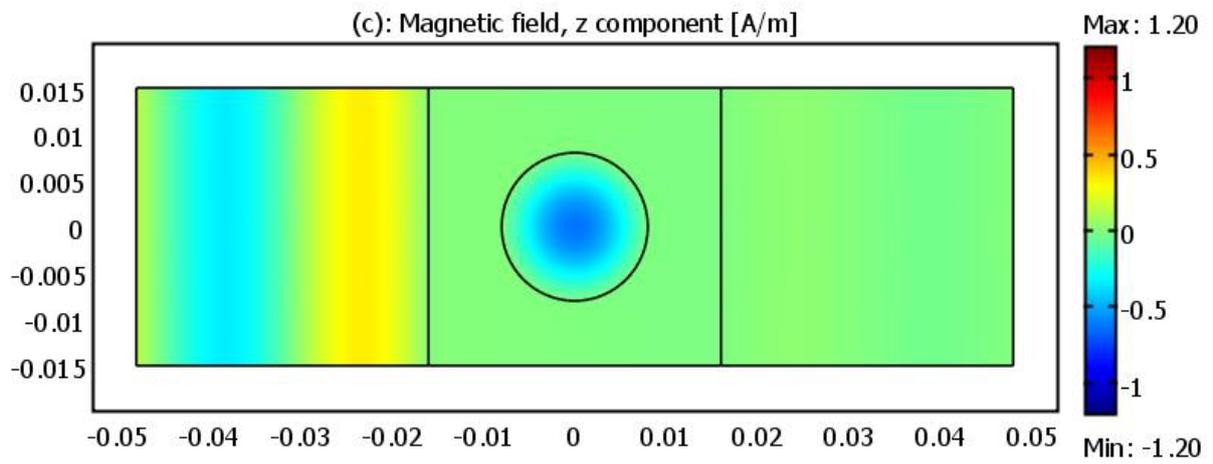

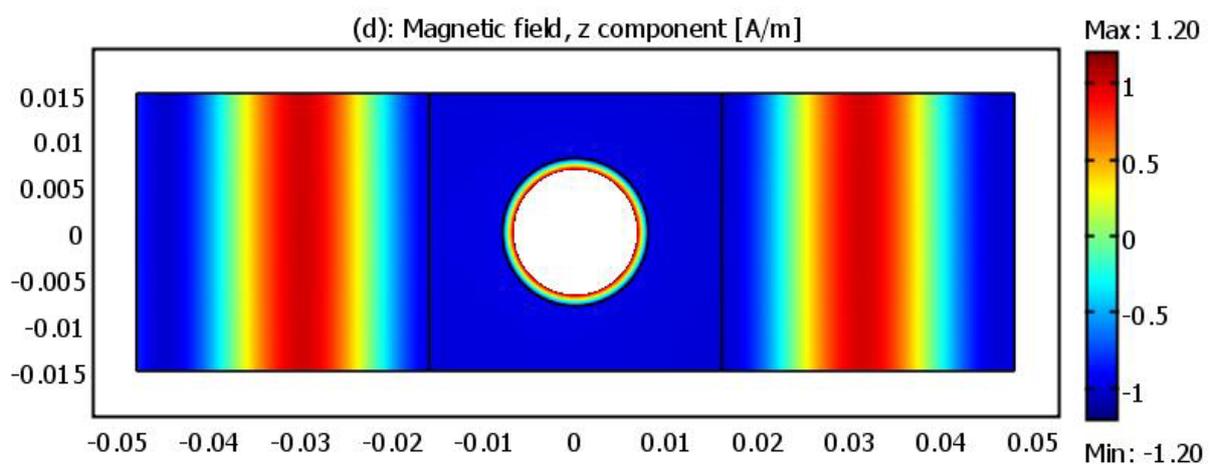

FIG. 2 (color outline). (a) The magnetic distribution of a MIZIM with $\epsilon_1 = 0.01$ and $\mu_1 = 0.01$. (b) The magnetic distribution of an ENZ medium with $\epsilon_1 = 0.01$ and $\mu_1 = 1$. (c) The magnetic distribution of an ENZ medium with a dielectric cylindrical defect with $\epsilon_2 = 2.06$ and $\mu_2 = 1$. (d) The magnetic distribution of an ENZ medium with a dielectric cylindrical defect with $\epsilon_2 = 2.405$ and $\mu_2 = 1$.

FIG. 2(a) shows the magnetic field distribution when a TM plane wave incident from the left impinges into a MIZIM with $\epsilon_1 = 0.01$ and $\mu_1 = 0.01$. The incoming TM wave transmits through the structure completely. If the permeability of the MIZIM is replaced by $\mu_1 = 1$, *i.e.* an ENZ medium, then only part of the incoming EM wave can transmit, as shown in FIG. 2(b). Now, if we embed a dielectric defect described above with $\epsilon_2 = 2.06$ and $\mu_2 = 1$ inside the ENZ medium, the structure will totally block the wave, as shown in FIG. 2(c). Likewise, the total transmission can be achieved for the same structure by modifying the material parameters of the cylindrical defects. The total transmission happens when $T$ is equal to 1. By substituting it into Eq. (10), we find we could simply replace the above dielectric defect with another one with $\epsilon_2 = 2.405$ and $\mu_2 = 1$ to fulfill the requirement. FIG. 2(d) shows that such a structure has a total transmission.

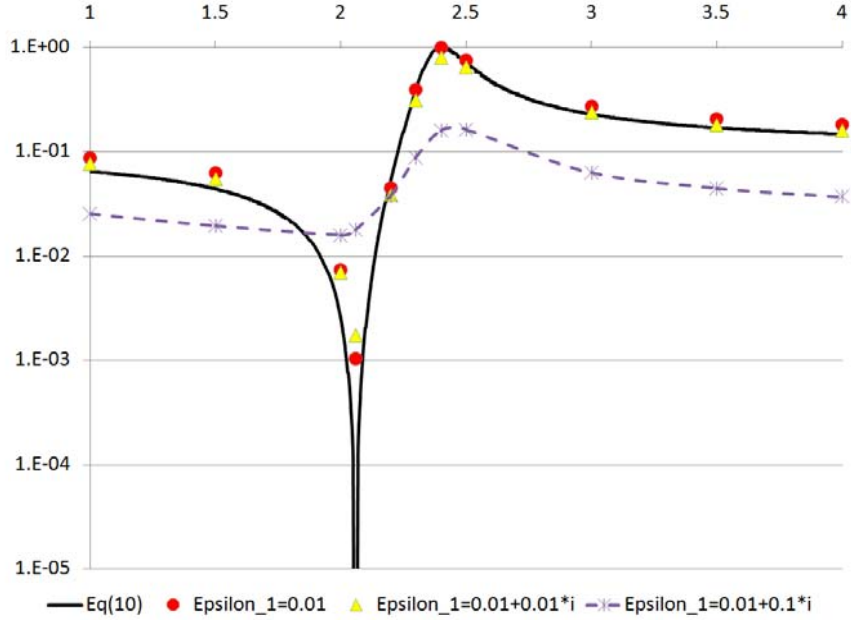

FIG. 3 (color online) The relationship between the transmission efficiency and the permittivity of the dielectric defect. The black solid curve denotes the results obtained directly from Eq. (10). The circular data is obtained from numerical simulations with an ENZ medium with $\epsilon_1 = 0.01$. The triangular data is obtained from numerical simulations with an ENZ medium with $\epsilon_1 = 0.01 + 0.01 * i$. The star-shaped data (together with the purple dashed curve) is from numerical simulations with an ENZ medium with $\epsilon_1 = 0.01 + 0.1 * i$.

As the above two dielectric defects have very close permittivities, we can consider using tunable refractive index materials (such as liquid crystals [19]) to control the wave transmission. In FIG. 3, we plot the relationship between the transmission efficiency and the permittivity of the dielectric defect (the black solid curve, $\epsilon_2$ ranges from 1 to 4). The curve shows that there is a total reflection at $\epsilon_2 = 2.06$ and a total transmission at $\epsilon_2 = 2.405$, as are already shown in FIG. 2(c) and (d), respectively. During the numerical simulations, we have used an ENZ medium with $\epsilon_1 = 0.01$ and $\mu_1 = 1$. Such an approximation is very accurate, as is also shown in FIG. 3 (see the circular data, which are very close to the results from Eq. (10)). In a real ENZ medium (either designed from metamaterials or using the plasmonic materials, such as metals), there should be some absorption. We consider two cases, one is an ENZ medium with $\epsilon_1 = 0.01 + 0.01 * i$, the other is with $\epsilon_1 = 0.01 + 0.1 * i$. The numerical results are shown in FIG. 3 by triangular data and star-shaped data (together with a purple dashed curve), respectively. A considerable amount of loss (such as the case of $\epsilon_1 = 0.01 + 0.1 * i$) will compromise both the total reflection and the total transmission effect. However, we can still use such a structure to control the wave transmission. The structure might be applied in on-chip applications such as switches and optical modulators.

In conclusion, we have found that a total reflection or transmission could be achieved by introducing suitable defects into an ENZ medium, instead of a MIZIM. Numerical simulations confirm our theory. Compared to the MIZIM, the ENZ medium is much easier to make with metamaterials or plasmonic materials, therefore gives possibilities to a feasible proof-of-principle experiment in future. Such a structure suggests promising on-chip applications.

## Acknowledgements

We thank Drs. Yun Lai and Xiaoping Hong for their helpful discussions.This work was supported by the National Natural Science Foundation of China under Grant No.11004147 and the Natural Science Foundation of Jiangsu Province under Grant No. BK2010211.

## References

[1] D. R. Smith, W. J. Padilla, D. C. Vier, S. C. Nemat-Nasser, and S. Schultz, Phys. Rev. Lett. **84**, 4184(2000).

[2] D. R. Smith, J. B. Pendry, and M. C. K. Wiltshire, Science **305**, 788 (2004).

[3] U. Leonhardt, Science **312**, 1777 (2006).

[4] J. B. Pendry, D. Schurig, and D. R. Smith, Science **312**, 1780 (2006).

[5] H. Y. Chen, C. T. Chan, and P. Sheng, Nature Mater.**9**, 387 (2010).

[6] V. G. Veselago, Sov. Phys. Usp. **10,** 509 (1968).

[7] H.-K. Yuan, U. K. Chettiar, W. Cai, A. V. Kildishev, A. Boltasseva, V. P. Drachev, and V. M. Shalaev, Opt. Express **15,** 1076 (2007).

[8] S. Enoch, G. Tayeb, P. Sabouroux, N. Guérin, and P. Vincent, Phys. Rev. Lett. **89,** 213902 (2002).

[9] R. W. Ziolkowski, Phys. Rev. E **70,** 046608 (2004).

[10] J. Li, L. Zhou, C. T. Chan, and P. Sheng, Phys. Rev. Lett. **90,** 083901 (2003).

[11] M. Silveirinha and N. Engheta, Phys. Rev. Lett. 97, 157403 (2006).

[12] M. Silveirinha and N. Engheta, Phys. Rev. B 75, 075119 (2007).

[13] M. Silveirinha and N. Engheta, Phys. Rev. B 76, 245109 (2007).

[14] A. Alù, M.Silveirinha, A. Salandrino, and N. Engheta, Phys. Rev. B **75,** 155410 (2007).

[15] R. Liu, Q. Cheng, T. Hand, J. J. Mock, T. J. Cui, S. A. Cummer, and D. R.Smith, Phys. Rev. Lett. **100,** 023903 (2008).

[16] B. Edwards, A. Alù, M. E. Young, M. Silveirinha, and N. Engheta, Phys.Rev. Lett. **100,** 033903 (2008).

[17] R. J. Pollard, A. Murphy, W. R. Hendren, P. R. Evans, R. Atkinson, G. A. Wurtz, A. V. Zayats, and V. A. Podolskiy, Phys. Rev. Lett. **102,** 127405 (2009).

[18] J.Hao, W. Yan, and M.Qiu, Appl. Phys. Lett. **96**, 101109 (2010).

[19] V. C. Nguyen, L. Chen, and K.Halterman, Phys. Rev. Lett. **105**, 233908 (2010).